\documentstyle[12pt]{article} 
\begin{document}

\begin{center}
{\Large \bf
The Asymptotics of Pion Charge  Form Factor.}\\[2mm]

{\large\em A.F.\ Krutov,$^a$ V.E.\
Troitsky\phantom{,}$^b$}\\[2mm]

{\small
$^a$Samara State University, 443011 Samara, Russia,\\
$^b$Nuclear Physics Institute, Moscow State University,
119899 Moscow, Russia}\\[5mm]

\end{center}

\begin{abstract}
The asymptotics of pion charge form factor is obtained in the
framework of relativistic Hamiltonian dynamics for the infinite
value limit of momentum transfer and zero value limit of
constituent--quark mass.  It is shown that this asymptotics is
the same as given by the perturbative QCD and is determined by
relativistic kinematics only, being independent on the
constituent quarks interaction in pion.  \end{abstract}
\vspace{10mm}

It is now well established that the description of the
electromagnetic structure of composite quark systems at low and
intermediate momentum transfers, i.e. in the region of so called
"soft" processes, needs nonperturbative approaches.
Moreover, in the investigation of
composite systems of light quarks (e.g., the pion) one
has to take into account relativistic effects, which give
essential contribution at low momentum transfers.

During last years the relativistic Hamiltonian dynamics
(RHD) is widely used in this nonperturbative region for
the investigation of composite quark systems (the foundation
of the approach, references to calculations and some
historical aspects can be found in the reviews
\cite{Pol89}-- \cite{Lev93}, see also
~\cite{BaK95} --~\cite{KrT97c}). The RHD can be formulated in
different ways. The most widely used
forms are: the instant form dynamics, the point form dynamics
and the light front dynamics. The main feature of RHD
{\it versus} field theories is the extraction  of the
finite number of the most important
degrees of freedom in each
concrete case. The establishment of the relation between RHD and
space--time field theoretical QCD description is a
principal and as yet unresolved problem.

Now it is widely believed that RHD and QCD complement each
other. The conventional (perturbative) QCD is considered to
give reliable predictions in high--energy region, while
RHD is considered to describe behavior at rather low energy.
However, the boundary between perturbative and nonperturbative
regions is not well defined.  For example, in the case of
exclusive processes, including the calculation of elastic form
factors for hadrons the boundaries of the
perturbarive regime are different in different calculations
\cite{IsL89}.

To discuss the relation between RHD and QCD it seems interesting
to compare the predictions of both theories in
some energy range. One of surely established QCD predictions
concerns the asymptotic behavior at
$Q^2\to\infty$ ($Q^2 = -q^2$, where $q$ is momentum transfer)
of the elastic form factors of composite quark systems.
This behavior was described by classic quark counting laws
(see ~\cite{Mat94} for the details). For the pion, up to
logarithmic corrections, the asymptotics has the form:

\begin{equation}
F_\pi(Q^2)\quad\sim\quad Q^{-2}\>.
\label{as-qcd}
\end{equation}

In the present paper we show that the asymptotics
(\ref{as-qcd})
can be obtained in the frame of RHD instant form
\cite{BaK95} if $Q^2\to\infty\>, \>M\to$0 (where $M$ is the
constituent quark mass), and that the asymptotic behavior
has universal character: it does not depend on the choose
of concrete interaction model for constituent quarks and
is defined by the relativistic kinematics of two--quark
system only.

The limits defined above are connected with the following
meaning of the physical behavior at large momentum transfers. At
$Q^2\to\infty$ the quarks become free and
their mass becomes equal to the initial one (current mass). For
the pion with light quarks this means in fact that $M\to$ 0. If
a dimension parameter $b$ defines the characteristic hadron
scale (the confinement scale), then the limits have the
following meaning:  $Q^2\gg b$, $M\ll b$.

In our approach the electromagnetic pion form factor is
represented in the form of a functional, generated by the so
called free two--particle form factor
\cite{BaK95}.  This free form factor describes the
electromagnetic properties of the system of two free particles,
and in the case of pion it defines the electromagnetic current
matrix element for the system of two free quarks. If the free
quarks have the quantum numbers corresponding to pion, then the
matrix element has the form
\cite{BaK95}:
$$
<\vec P,\sqrt s,J,l,S,m_J |\,j_\mu\,| \vec
     P',\sqrt{s'} ,J',l',S',{m_J}'> =
$$
\begin{equation} =
A_\mu (s,Q^2,s')\> g_0 (s,Q^2,s')\>.
\label{param}
\end{equation}
Here $Q^2 = -(P - P')^2$,  $s = P^2$ -- system invariant mass
square,
$J = S = l =$0, $g_0 (s,Q^2,s')$ -- the free two--particle form
factor for the system of
particles without interaction. The vector $A_\mu (s,Q^2,s')$
is determined by the current transformation properties
(Lorentz--covariance and current conservation law)
\cite{BaK95}.  The explicit form of the function
$g_0(s,Q^2,s')$ in the case of point quarks was obtained in
\cite{KrT93}:

\begin{eqnarray}
g_0(s,Q^2,s')=
  \frac{(s+s'+Q^2)Q^2}{2\sqrt{(s-4M^2)
(s'-4M^2)}}\>
\frac{\theta(s,Q^2,s')}{{[\lambda(s,-Q^2,s')]}^{3/2}} (1+Q^2/4M^2)^{-1/2}\nonumber\\
\left\{(s+s' +Q^2)
\cos{(\omega_1+\omega_2)}+
  \frac{1}{M}\xi(s,Q^2,s')
\>
\sin{(\omega_1+\omega_2)} \right\}
\label{ff-nonint}
\end{eqnarray}
Here $\lambda(a,b,c)=a^2+b^2+c^2-2(ab+ac+bc),
\xi=\sqrt{ss'Q^2-M^2\lambda(s,-Q^2,s')}$,\\
$\omega_1$ ¨ $\omega_2$ - the parameters of Wigner rotation:
\begin{eqnarray}
\omega_1=\arctan{\frac{\xi(s,Q^2,s')}{M[(\sqrt s+\sqrt
{s'})^2+Q^2]+\sqrt{ss'}(\sqrt s+\sqrt{s'})}},\nonumber\\
\omega_2=\arctan{\frac{(2M+\sqrt s+\sqrt
{s'})\xi(s,Q^2,s')}{M(s+s'+Q^2)(2M+\sqrt s+\sqrt
{s'})+\sqrt{ss'}(4M^2+Q^2)}}
\label{Wign-param}
\end{eqnarray}
 $\theta(s,Q^2,s')=
\vartheta(s'-s_1)-\vartheta(s'-s_2)$, $\vartheta$ - the step
function,
\begin{displaymath} s_{1,2}=2M^2+\frac{1}{2M^2}
(2M^2+Q^2)(s-2M^2)\mp \frac{1}{2M^2}
\sqrt{Q^2(Q^2+4M^2)s(s-4M^2).}
\end{displaymath}
Free two--particle form factors were first introduced in
\cite{TrS69} and were used in
~\cite{KoT72} -- ~\cite{KiT75} for the description of composite
systems in the frame of the modified dispersion approach.
In fact, the form factor $g_0(s,Q^2,s')$ in (\ref{param})
is a regular distribution (generalized function) and is a linear
continuous functional on a test functions space \cite{Vla76}.

To be used for pion form factor the functional given by
$g_0 (s,Q^2,s')$ is to be defined on the space  of functions
$\varphi(s,s') = \psi(s)\,\psi(s')$
using the following prescription \cite{BaK95}:

\begin{equation}
F_\pi (Q^2)=\int d\sqrt s\ d\sqrt{s'}\ \psi (s)\,g_0(s,Q^2,s')\,
\psi (s')
\label{ff}
\end{equation}
$\psi$ are functions normalized with the relativistic
density of states:
$\psi(s) = \sqrt[4]{s}\,u(k)\,k$;  $u(k)$ are the two--quark
wave functions of the relative motion $S$ -- state,
$s = 4(k^2+M^2)$.

So, to obtain the asymptotic behavior of the pion form factor
one has to evaluate the asymptotics at
$Q^2\to\infty\>,\>M\to$0
of the generalized function
$g_0(s,Q^2,s')$, this function defined on a certain class of
functions giving the pion form factor.

The constituent--quark mass $M$ and the momentum--transfer
square $Q^2$ are the parameters of generalized function
$g_0(s,Q^2,s')$. So the limits
$Q^2\to\infty$ and $M\to$ 0 are to be considered in the
generalized sense (in the sense of distributions) as the limits
of generalized functional.  The possibility of realizing the
calculation of this limit for the generalized function
$g_0(s,Q^2,s')$  can be justified physically by the fact that
the pion in the asymptotic region can be considered as a free
quark system with pion quantum numbers and the electromagnetic
properties of such system are defined just by the free
two--particle form factor $g_0(s,Q^2,s')$.

To obtain the asymptotic expansion of the generalized  function we
shall make the following assumptions.

1. Let the generalized  function be given on the
space $R^2$. During the calculation of
(\ref{ff}) we have not used such assumption, in
(\ref{ff}) the variables
$s\>,\>s'$ were in the physical region
$s\>,\>s'\ge\>4M^2$.

2. For the test function space let us take as usually the
space ${\cal S}(R^2)$ of infinitely differentiable functions,
decreasing at $|\vec s|\to \infty$ ($\vec s = (s,s')$) with all
its derivatives faster than any degree of
1/$|\vec s|$ \cite{Vla76}. This means that test
functions decrease not slowlier than
$\exp(-\,\beta\,|\vec s|^\alpha)\>,\alpha>$ 0.

3. Let us define the functional giving the regular generalized
function by
\begin{equation}
<\tilde g_0 (s,Q^2,s')\,,\varphi(s,s')> = \int\,d\mu(s,s')\,
\tilde g_0 (s,Q^2,s')\,\varphi(s,s')\>.\label{<>}
\end{equation}
Here
\begin{equation}
\tilde g_0 (s,Q^2,s') = 16\,\vartheta(s - 4M^2)\,
\vartheta(s' - 4M^2)\,g_0 (s,Q^2,s')\label{tilde g0}
\end{equation}
$$
d\mu(s,s') = \sqrt[4]{ss'}\,d\mu(s)\,d\mu(s')\>,\quad d\mu(s) = \frac{1}{4}
k\,d\sqrt{s}\>.
$$
$\vartheta(x)$ -- the standard step function. If
$\varphi(s,s') = u(k(s))\,u(k'(s'))$
then the functional (\ref{<>}) has the form
(\ref{ff}).

In fact the calculation of the
asymptotics of the generalized  function is just the calculation
of the asymptotics of the functional (\ref{<>}) when
$Q^2\to\infty\>,\quad M\to$0.

To estimate the asymptotics of the integral in
(\ref{<>}) let us realize the following steps:

1. Let us transform the integral in $s'$ to the integral
on the segment [0,1]. To perform this transformation we
use the fact that the free two--particle form factor
$g_0 (s,Q^2,s')$ contains the cutting functions
$$
  \vartheta(s' - s_1(s,Q^2)) - \vartheta(s' - s_2(s,Q^2))\>.
$$

2. Let us introduce the function
$\tilde \varphi (s,s')$  which is monotone
decreasing function of the variable $s$ when $s'$ is fixed
and of $s'$ when $s$ is fixed ($\>s,s'>$0) and such that
\begin{equation} |\varphi(s,s') - \tilde
\varphi(s,s')|\to 0\>,\quad |\vec s| \to \infty\>\label{tilde
phi} \end{equation}

3. Let us introduce two dimensionless parameters and let us
realize the limiting procedure keeping
$\eta = Q^2/b^2\to\infty$, $\xi = b/2M\to\infty$.

4. In the integral in $s$ let us introduce the new variable
$y=E/2M\>, s= 4M^2 + 2ME$, and let us separate explicitly
the dependence on the parameter $\eta$ introduced above.

5. Let us restrict ourselves by the main term of the
integrand in the functional (\ref{<>}) when
$\eta\to\infty$. This can be done because
the integrals do converge uniformly.

As the result we obtain the following estimate for the
functional (\ref{<>}) when
$\eta\to\infty$:
$$ <\tilde g_0
(s,Q^2,s')\,,\varphi(s,s')>\> \sim
\>\frac{1}{16}\sqrt[4]{\eta}\,
\left(\frac{2M}{b}\right)^{5/2}b^3\,\int\limits_0^\infty\,dy
\frac{\sqrt{y}}{(1 + y)^{1/4}}\cdot
$$
$$
\cdot\>\int\limits_0^1\,dt\,\frac{\tilde\varphi(s,\tilde s)}
{a(t)^
{1/4}}\,\frac{1}{\left[\sqrt{y}(2t - 1) + \sqrt{1 + y}\right]^2}\>\cdot
$$
\begin{equation}
\cdot\left\{\left[\sqrt{y}(2t - 1) + \sqrt{1 + y}\right]\,
\cos(\tilde \omega_1 +
\tilde \omega_2)
+ 2\sqrt{y}\sqrt{t(1-t)}\sin(\tilde \omega_1 +
\tilde \omega_2)\right\}\>.\label{gl eta}
\end{equation}
In (\ref{gl eta}): $\tilde \omega_{1,2}$ -- are the main terms
in  the expansion of the spin rotation parameters which enter
$g_0 (s,Q^2,s')$ \cite{BaK95}, $$ \tilde s = \eta\,b^2\,a(t) =
\eta\,b^2\left(1 + \frac{s - 4M^2}{2M^2} +\frac{1}{2M^2}\,
(2t - 1)\,\sqrt{s(s - 4M^2)}\right)\>.
$$

To estimate the inner integral in
(\ref{gl eta}) let us separate explicitly the parameter
$\xi$, introduced above and let us obtain the asymptotics of
the integral when $\xi\to\infty$.
The main term has the following form:
\begin{equation}
J(\eta\,,\xi\,,y)\>\sim\>\sqrt{\frac{b}{2\xi}}\,\frac{1}{\sqrt[4]{s - 4M^2}}\,
\int\limits_0^1\,dt\,\tilde\varphi(s,\tilde s_0(t))\,f(t,y)\>,\label{J}
\end{equation}
\begin{equation}
f(t,y) = \frac{t^{1/4}}{
\left[\sqrt{y}(2t - 1) + \sqrt{1 + y}\right]^2}
\cdot\left\{\sqrt{y}(2t - 1) + \sqrt{1 + y}
+ 2\sqrt{y}{(1-t)}
\right\}\>.\label{gl ksi}
\end{equation}
Here
\begin{equation}
\tilde s_0(t) = 4\,\eta\,\xi^2\,(s - 4M^2)\,\left(1 + \frac{b^2}{2(s - 4M^2)
\,\xi^2}\right)\,t\>.\label{s0}
\end{equation}
To obtain
(\ref{gl ksi}) we have used the following forms valid when
\\ $\xi\to\infty$:
$$
\cos(\tilde \omega_1 +\tilde \omega_2)\>\sim\>\sqrt{t}\>,\quad
\sin(\tilde \omega_1 +\tilde \omega_2)\>\sim\>\sqrt{1-t}\>.
$$
Let us show that
it is a small $\varepsilon$--neighborhood
of the point $t=0$ that gives the main contribution to the
asymptotic expansion of the integral
(\ref{gl ksi})
(see \cite{Olv90}).
To do this let us estimate the value at
$\eta\>,\xi\>\to \infty$
of the following integral:
\begin{equation}
\int\limits_\varepsilon^1\,dt\,\tilde\varphi(s,\tilde s_0(t))\,f(t,y)\>,
\end{equation}
where $\varepsilon$
has a small but fined value.
The function  $\tilde\varphi(s,\tilde s_0(t))$ is a
monotone decreasing function of
the variable $t$
because of
(\ref{s0}) and thus
$$ \left|\tilde\varphi(s,\tilde s_0(t))\right|\>\le\>
\left|\tilde\varphi(s,\tilde s_0(\varepsilon))\right|\>.
$$
So the following is valid:
$$
\left|\int\limits_\varepsilon^1\,dt\,\tilde\varphi(s,\tilde s_0(t))\,f(t,y)
\right|\>\le\>
\int\limits_\varepsilon^1\,dt\,\left|\tilde\varphi(s,\tilde s_0(t))
\right|\,\left|f(t,y)\right|\>\le
$$
\begin{equation}
\le\>\left|\tilde\varphi(s,\tilde s_0(\varepsilon))\right|\,
\int\limits_\varepsilon^1\,dt\,\left|f(t,y)\right|\>.\label{est}
\end{equation}
The last integral in
(\ref{est}) converges
if the function $f(t,y)$ is given by
(\ref{gl ksi}).
It follows from (\ref{tilde phi}) that the rate of decreasing of
the function $\left|\tilde\varphi(s,\tilde
s_0(\varepsilon))\right|$
is stronger than any inverse power of
$\tilde s_0(\varepsilon)$, or, by
(\ref{s0}), stronger than any inverse power of
$\eta$ and $\xi^2$.  So, the main contribution
to the asymptotics of the integral
(\ref{J})  is given by a small neighborhood
of the point $t=0$ and is of power law.

Let us now write the following expansion in the
$\varepsilon$--vicinity of the point $t=0$:
$$
\tilde \varphi(s,\tilde s_0)\>\simeq\>\tilde \varphi(s,0) +
4\,\eta\,\xi^2\,(s - 4M^2)\left(1 + \frac{1}{2}\,\frac{b^2}{s - 4M^2}
\frac{1}{\xi^2}\right)\,\tilde\varphi'_{s'}(s,0)\,t\>
$$
After simple transformations we now obtain the following form
for the asymptotics of the integral in
(\ref{J}):
$$
J(\eta\,,\xi\,,y)\>\sim\>\sqrt{\frac{b}{2\xi}}\,\frac{1}{\sqrt[4]{s - 4M^2}}\,
\tilde\varphi(s,0)\,\frac{\sqrt{y +1} + \sqrt{y}}
{(\sqrt{y + 1} - \sqrt{y})^2}\cdot
$$
\begin{equation}
\cdot\int\limits_0^\varepsilon\,dt\,t^{1/4}\,\exp\left[-\,4\eta\xi^2\,
(s - 4M^2)\left(1 + \frac{1}{2}\,\frac{b^2}{s - 4M^2}\frac{1}{\xi^2}
\right)\left|\frac{\tilde \varphi'_{s'}(s,0)}
{\tilde \varphi(s,0)}\right|t
\right]\>.\label{int e}
\end{equation}
The appearance of the absolute value in
(\ref{int e}) is the consequence of the monotonity of the
function $\tilde\varphi(s\,,\,s')$.
Let us use the new variable in
(\ref{int e}):
$$ v = 4\,\xi^2\, (s - 4M^2)\left(1 +
\frac{1}{2}\,\frac{b^2}{s - 4M^2}\frac{1}{\xi^2}
\right)\left|\frac{\tilde \varphi'_{s'}(s,0)}{\tilde \varphi(s,0)}\right|t\>.
$$
and take into account the fact that to within exponentially
decreasing terms the following estimate is valid
\cite{Olv90}:
$$
\int\limits_0^{\varepsilon_v}\,dv v^{1/4}\hbox{e}^{-\eta\,v} \simeq
\int\limits_0^\infty\,dv v^{1/4}\hbox{e}^{-\eta\,v} = \frac{1}{\eta^{5/4}}\,
\Gamma\left(\frac{5}{4}\right)\>.
$$
Here $\Gamma(x)$ is Euler gamma-function.

Thus, we have the following asymptotic estimate for the integral
(\ref{int e}):
$$
J(\eta\,,\xi\,,y)\>\sim\>\sqrt{\frac{b}{2\xi}}\,\frac{1}{\sqrt[4]{s - 4M^2}}\,
\tilde\varphi(s,0)\,\frac{\sqrt{y +1} + \sqrt{y}}
{(\sqrt{y + 1} - \sqrt{y})^2}\cdot
$$
$$
\cdot\frac{1}{(2\xi)^{5/2}}
\left[\frac{\tilde \varphi(s,0)}{\left|\tilde \varphi'_{s'}(s,0)\right|}
\right]^{5/4}\frac{\Gamma\left(\frac{5}{4}\right)}{(s - 2M^2)^{5/4}}
\frac{1}{\eta^{5/4}}\>.\label{in as}
$$
Let us now substitute
this estimate in Eq. (\ref{gl eta}) and return to the $s$ --
integration:
$$ <\tilde g_0 (s,Q^2,s')\,,\varphi(s,s')>\> \sim
\frac{1}{4}\frac{M^4}{Q^2}\,
\Gamma\left(\frac{5}{4}\right)\,\int\,d\sqrt{s}\,\,
\frac{\sqrt[4]{s(s - 4M^2)}}{(s - 2M^2)^{5/4}}\cdot
$$
\begin{equation}
\cdot\frac{\sqrt{s} + \sqrt{s - 4M^2}}{(\sqrt{s} - \sqrt{s - 4M^2})^2}
\left|\frac{\tilde \varphi(s,0)}{\tilde\varphi'_{s'}(s,0)}\right|^{5/4}
\tilde\varphi(s,0)\>.\label{fin}
\end{equation}

Up to a preasymptotic factor one can write
(\ref{fin}) in the form:
\begin{equation}
<\tilde g_0 (s,Q^2,s')\,,\varphi(s,s')>\quad \sim \quad
<\frac{1}{Q^2}\,g(s,s')\,,
\varphi(s,s')>\>.\label{fin fun}
\end{equation}
In the Eq.(\ref{fin fun}) the function $g(s,s')$  is given
by:
 $$
g(s,s') = M^4\,\frac{1}{\sqrt[4]{s(s - 4M^2)}}\frac{1}{(s -
2M^2)^{5/4}} \frac{\sqrt{s} + \sqrt{s - 4M^2}}{(\sqrt{s} -
\sqrt{s - 4M^2})^2}
\cdot $$ $$
\cdot\vartheta(s - 4M^2)\,\delta(\mu(s') - \mu(s))\>.\label{g(s,s')}
$$

Finally, we proved that in the sense of distributions
the following asymptotic estimate is valid:
\begin{equation}
\tilde g_0(s,Q^2,s')\>\sim\>\frac{1}{Q^2}\,g(s,s')\>,\quad
Q^2\>\to\infty\>, \quad M\>\to\>0\>.\label{g0-g} \end{equation}

Let us note that this result is also valid
in the case when the test function space
is larger than ${\cal S}(R^2)$. It is
necessary only
that the test functions ensure the uniform convergence of the
integrals entering the functional
(\ref{<>}).

Applying the result (\ref{g0-g})
to the Eq.(\ref{ff}), one can see that the asympotics
does not depend on the choose of the functions
$u(k)$, that is on the concrete form of the quark interaction in
pion.

Thus, we obtained the pion form factor
asymptotics which
1) coincides with QCD asymptotics
(\ref{as-qcd}), 2) does not depend on the interaction model for
quarks in pion, 3) is defined by the free two--particle form
factor considered as generalized function. This free
two--particle form factor is obtained through relativistic
kinematics method, so the asymptotics is
due to the relativistic kinematics.

Let us compare our result with the results obtained by
other approaches. The asymptotic behavior of the
form factor of nucleon considered as pion -- nucleon bound
state in $P_{11}$ -- channel was considered in \cite{KiS72}.
The asymptotics of the electromagnetic form
factor was obtained in terms of the asymptotics of the
$\delta_{11}$ phase shift of elastic  $\pi\,N$ --
scattering. The system, as in the our case, contains two
constituents. However, those constituents are real particles
with spins 0 and 1/2, so that two results can not be compared
directly. Nevertheless, there is some kind of similarity
of results. The formal limit $m_\pi\>,\>m_N\to$ 0
gives the asymptotics of electromagnetic form factor of
the composite system ~\cite{KiS72} which does not depend
on the character of constituents interaction: the asymptotics
ceases to depend on the pion--nucleon scattering phase shift.

In the frame of light front RHD
\cite{Ter76} the QCD type asymptotic behavior of pion form
factor when
$Q^2\to\infty\>,\> M\to$0 was obtained in the case of special
power form restriction for the wave function asymptotics.
In papers \cite{ChC88}, \cite{Kei94} the asymptotics of the form
(\ref{as-qcd}) was found for the gaussian wave function.

There is a principal difference between our results and the
results of above mentioned papers. In those papers the results
are the cosequence of the structure of the argument of the wave
function for interacting quarks and of the special limitations
for wave function asymptotics. It seems difficult to interprete
this fact from the asymptotic freedom point of view which in
fact is the base of well known result (\ref{as-qcd}).
So the behavior of the two--quark system form factor
in the range $Q^2\to\infty$, where quarks are free, appears
to be a kind of "memory" about the quark interaction.

In our approach the result coinciding with (\ref{as-qcd})
follows directly from relativistic kinematics
which is valid for all energies.  Let us emphasize
that if we neglect, for example, the relativistic spin rotation
effect  (if we let the rotation parameters $\omega_{1,2}$ in
$g_0(s,Q^2,s')$ be zero) then the asymptotics of the generalized
function in (\ref{g0-g}) is immediately changed. This is in
contrast with the case of light front dynamics. It is easy to
show that the analog of the relativistic spin rotation in the
light front dynamics -- Melosh rotation \cite{ChC88} -- does
not affect the pion form factor asymptotics.

Let us give now the results for pion form factor asymptotics in
the cases of harmonic oscillator model
\cite{ChC88}:
\begin{equation}
u(k) = N_{HO}\exp(-k^2/2b^2)\>.\label{HO}
\end{equation}
and for power law wave function
\cite{CaG94}, \cite{Sch94b}:
\begin{equation}
u(k) =N_{PL}\,{(k^2/b^2 +
1)^{-n}}\>.\label{PL}
\end{equation}

The gaussian form of wave function makes the problem of the
calculation of asymptotics much simplier because in this case
one can use the theorems from \cite{Olv90}.

We use the model (\ref{HO}) to demonstrate the role of
relativism and, in particular, the role of the relativistic spin
rotation effect. The estimate for the pion form factor including
relativistic spin rotation effect gives
\begin{equation}
F_\pi(Q^2)\>\sim 32\sqrt{2}\frac{b^2}{Q^2}\>,\label{as s pov}
\end{equation}
while without relativistic spin rotation:
\begin{equation}
F_\pi(Q^2)\>\sim\> 4\sqrt{2}\frac{M}{Q}\>.  \label{as bez pov}
\end{equation}
The contribution of the relativistic spin rotation is
essential. If we neglect this contribution,
we do not obtain correct (i.e.coinciding with QCD) asymptotics.

To clear up the role of relativism in our calculation
let us write the nonrelativistic asymptotics obtained from
the asymptotic estimate of nonrelativistic limit of Eq.
(\ref{ff}) in the case of the model (\ref{HO}):
\begin{equation}
F_\pi(Q^2)\sim\>2\sqrt{2}\sqrt{\frac{Q}{2M}}\exp\left(-\,\frac{Q^2}{8b^2}
\right)\>.
\label{as ner}
\end{equation}
The nonrelativistic asymptotics gives Gaussian decrease for the
form factor. The comparison of
(\ref{as s pov}) and (\ref{as bez pov}) with (\ref{as ner})
emphasizes the role of relativism. Let us note, that in
(\ref{as s pov})-- (\ref{as ner})
the explicit form of the normalization constant in
(\ref{HO}) is taken into account:
$N_{HO} = (4\,/\sqrt\pi\,b^3)^{1/2}$.

In the case of the model
(\ref{PL}) the form--factor asymptotics has the form:
\begin{equation}
F_\pi(Q^2) \>\sim\> 2\frac{b^2}{Q^2}\,\left(b^3\,N_{PL}^2\right)
\frac{\Gamma\left(\frac{5}{4}\right)}{n^{5/4}}\,
B\left(\frac{5}{4}\,,\,n - \frac{5}{4}\right)\>,\quad n\>>\>\frac{5}{4}
\label{n}
\end{equation}
Here $B(x\,,\,y)$ is Euler beta--function. Note, that  (\ref{n})
shows that the asymptotic (in the sense of distributions) form
(\ref{g0-g}) is really valid for the test function space larger
than ${\cal S}(R^2)$.

Let us note, that the asymptotic forms
(\ref{as s pov}) and (\ref{n}) do not contain the dimension
parameter of quark mass. This fact is connected with the
physics of asymptotically large momentum transfers when
all scales are determined by the confinement scale.
In the models (\ref{HO}) and (\ref{PL}) it is $b$ that plays
in fact the role of such parameter.

To conclude, let us formulate our main results.
1) In the frame of instant form relativistic
Hamiltonian dynamics the asymptotics  at $Q^2\to\infty$, $M\to$0
of the pion form factor coincides with that given by
perturbative QCD.
2) The asymptotics is determined only by the relativistic
kinematics, in particular, by the relativistic spin rotation
effect, and does not depend on the choose of quark wave
function in pion, i.e. on the model for quark interaction. \\[5mm]

The authors thank A.I.Kirillov for helpfull discussion and
valuable comments.

The work is supported in part by the Russian Foundation for
Basic Researches, Grant No. 96 -- 02 -- 17288.\\[5mm]

\end{document}